\documentclass[12pt]{article}
\usepackage{graphicx}
\def\gtrsim{\mathrel{\hbox{\rlap{\hbox{\lower4pt\hbox{$\sim$}}}\hbox{$>$}}}}
\def\lesssim{\mathrel{\hbox{\rlap{\hbox{\lower4pt\hbox{$\sim$}}}\hbox{$<$}}}}
\begin{document}

\begin{center}
{\LARGE \bf 
Updated AGASA event list above 4 $\times$ 10$^{19}$eV
}
\end{center}
%%\vspace{0.5zh}

\begin{center}
{N. Hayashida$^{1}$,} 
{K. Honda$^{2}$,}
{N. Inoue$^{3}$,}
{K. Kadota$^{4}$,}
{F. Kakimoto$^{4}$,}
{S. Kakizawa$^{5}$,}		\\
{K. Kamata$^{6}$,}
{S. Kawaguchi$^{7}$,}
{Y. Kawasaki$^{8}$,}
{N. Kawasumi$^{9}$,}
{E. Kusano$^{10}$,}		\\
{A. M. Mahrous$^{3}$,}
{K. Mase$^{1}$,}
{T. Minagawa$^{1}$,}
{M. Nagano$^{11}$,}
{D. Nishikawa$^{1}$,}
{H. Ohoka$^{1}$,}		\\
{S. Osone$^{1}$,}
{N. Sakaki$^{1}$,}
{M. Sasaki$^{1}$,}
{K. Shinozaki$^{3}$,}
{M. Takeda$^{1}$,}
{M. Teshima$^{1}$,}		\\
{R. Torii$^{1}$,}
{I. Tsushima$^{9}$,}
{Y. Uchihori$^{12}$,}
{T. Yamamoto$^{1}$,}
{S. Yoshida$^{1}$,}
{and H. Yoshii$^{13}$}
\end{center}

\begin{center}
\begin{small}
$^{1}$
Institute for Cosmic Ray Research, University of Tokyo, Tokyo 188-8502, Japan

$^{2}$
Faculty of Engineering, Yamanashi University, Kofu 400-8511, Japan

$^{3}$
Department of Physics, Saitama University, Urawa 338-8570, Japan

$^{4}$
Department of Physics, Tokyo Institute of Technology, Tokyo 152-8551, Japan

$^{5}$
Faculty of Science, Shinshu University, Matsumoto 390-8621, Japan

$^{6}$
Nishina Memorial Foundation, Komagome, Tokyo 113-0021, Japan

$^{7}$
Faculty of General Education, Hirosaki University, Hirosaki 036-8560, Japan

$^{8}$
RIKEN (The Institute of Physical and Chemical Research), Saitama 351-0198, Japan

$^{9}$
Faculty of Education, Yamanashi University, Kofu 400-8510, Japan

$^{10}$
KEK, High Energy Accelerator Research Organization, Institute of Particle and Nuclear Studies, Tsukuba 305-0801, Japan\\

$^{11}$
Department of Applied Physics and Chemistry, Fukui Institute of Technology, 
Fukui 910-8505, Japan 

$^{12}$
National Institute of Radiological Sciences, Chiba 263-8555, Japan

$^{13}$
Department of Physics, Ehime University, Matsuyama 790-8577, Japan
\end{small}
\end{center}

%*********************************************************************
%%\vspace{1zh}
\begin{abstract}
After our Ap.J. publication of the Akeno Giant Air Shower Array (AGASA) 
results in 1999 (Takeda et al., 1999), 
we observed nine events with energies above 4 $\times$ 10$^{19}$eV 
until May 2000.
This short report lists the coordinates of these events, and 
shows the updated energy spectrum and arrival direction map. 
The analysis was carried out with the same procedure employed in the
Ap.J. publication.
\end{abstract}

%%%%%%%%%%%%%%%%%%%%%%%%%%%%%%%%%%%%%%%%%%%%%%%%%%%%%%%%%%%%%%%%%%%%%%%%%%%%%%%
\section{Introduction}
\label{sect:intro}

The Akeno Giant Air Shower Array (AGASA) has been operated 
in stable since 1990 and 
the exposure exceeds 4.0 $\times$ 10$^{16}$ m$^2$ s sr 
until the end of May 2000.
Since there are many requests to use new events after the publication of
the event list in PRL \cite{takeda98a} and Ap.J. \cite{takeda99a}, 
we show the updated list in this report. 
The analysis was carried out with the same
parameters used in the previous papers \cite{takeda98a,takeda99a}.  

Although the zenith angle range is limited less than 
45$^\circ$ in the present analysis, 
our dataset has now enough statistics to determine
experimentally the lateral distribution of shower particles and
the attenuation of S(600) with atmospheric depth 
to larger zenith angles in energy range of $\gtrsim$ 3 $\times$ 10$^{19}$eV. 
Our energy estimation procedure for larger zenith angles is now 
under careful investigation, 
and we may be able to apply a refined attenuation length beyond 45$^\circ$. 
This may modify the parameters in the energy estimation procedure 
slightly; but 
we can extend our observable sky to larger zenith angles than 45$^\circ$
and increase the effective exposure of AGASA
in our next full-paper publication. 

\newpage
In this report, 
we first show the updated energy spectrum above 3 $\times$ 10$^{18}$eV 
with events until May 2000. 
And then we show the cosmic-ray arrival direction distribution 
with the additional dataset 
of the Akeno 20 km$^{2}$ array (A20) before 1990. 
Important informations for the present analysis are listed below, 
and for details refer our publications 
\cite{takeda98a,takeda99a,teshima86a,dai88a,chiba92a,ohoka97a,yoshida94a,yoshida95a}. 
\begin{flushleft}
\begin{tabular}{p{0.03\textwidth}ll}
& Exposure :			& 4.0 $\times$ 10$^{16}$ m$^{2}$ s sr	\\
& Energy conversion formula :	& 
	E $=$ 2.03 $\times$ 10$^{17}$ S$_{0}$(600) \hspace{0.8em} eV	\\
& Attenuation of S(600) :		&
	$ S_{\theta}(600) = S_0(600) \exp \left[ 
	- \frac{920}{500} \left( \sec \theta - 1 \right) 
	- \frac{920}{594} \left( \sec \theta - 1 \right)^2 \right] $	\\
\end{tabular}
\begin{tabular}{p{0.03\textwidth}ll}
& Error in S(600) determination :	& $\pm$ 30 \% above 10$^{19}$eV	\\
& Error in arrival direction determination : 
				& 1.6$^\circ$ above 4 $\times$ 10$^{19}$eV \\
\end{tabular}
\end{flushleft}

\begin{figure}[ht]
%%\vspace{1zh}
\centerline{\includegraphics[,width=0.795\textwidth]{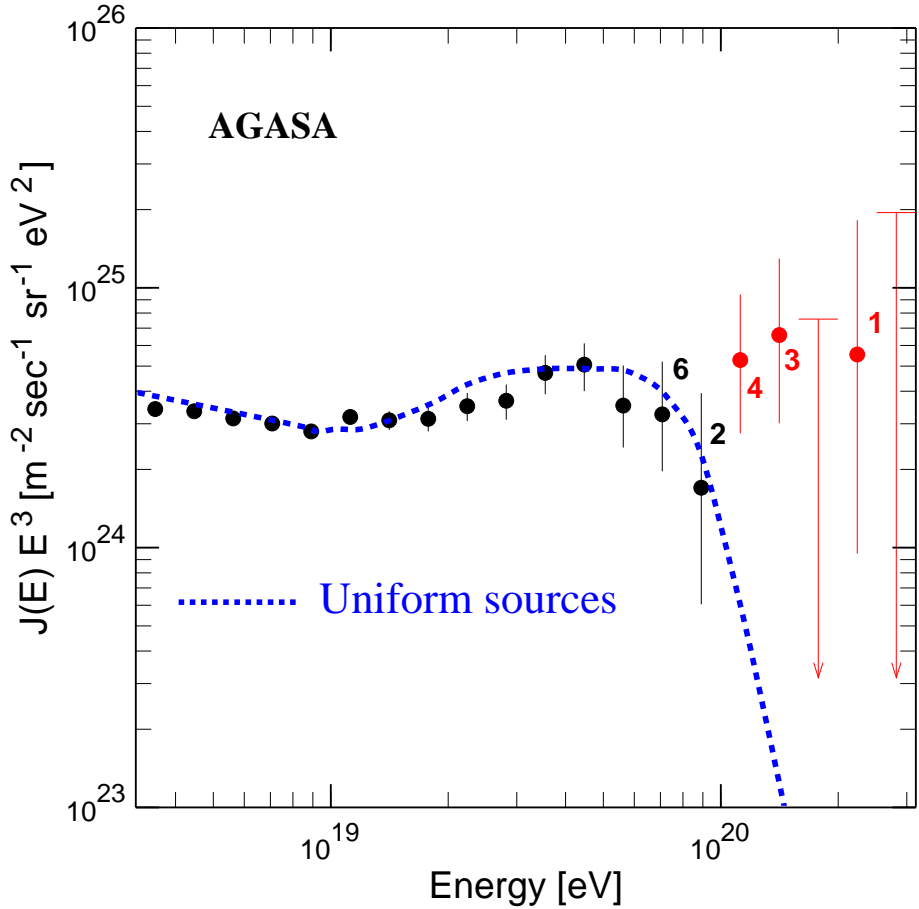}}
\caption{Energy spectrum observed with AGASA.
	The vertical axis is multiplied by $ E^{3} $. 
	Error bars represent the Poisson upper and lower limits 
	at $ 68 \% $ and arrows are $ 90 \% $ C.L. upper limits. 
	Numbers attached to points show the number of events 
	in each energy bin.
	The dashed curve represents the spectrum expected for 
	extragalactic sources distributed uniformly in the Universe, 
	taking account of the energy determination error.
	\label{fig:spectrum}
}
\end{figure}
%

%*********************************************************************
\section{Updated Results}
\label{sec:results}

The updated energy spectrum observed with AGASA (without A20)
is shown in Figure \ref{fig:spectrum}, 
multiplied by E$^{3}$ in order to emphasize details of 
the steeply falling spectrum. 
Error bars represent the Poisson upper and lower limits at $ 68 \% $
and arrows are $ 90 \% $ C.L. upper limits. 
Numbers attached to points show the number of events in each energy bin. 
The dashed curve represents the spectrum expected for 
extragalactic sources distributed uniformly in the Universe, 
taking account of the energy determination error \cite{yoshida93a}. 
Now we observed 8 events above 10$^{20}$eV. 
% this is the evidence of extension of the cosmic-ray energy spectrum 
% beyond the predicted GZK cutoff \cite{g66a, zk66a}. 

\begin{figure}[ht]
%%\vspace{2zh}
\centerline{\includegraphics[,width=0.9\textwidth]{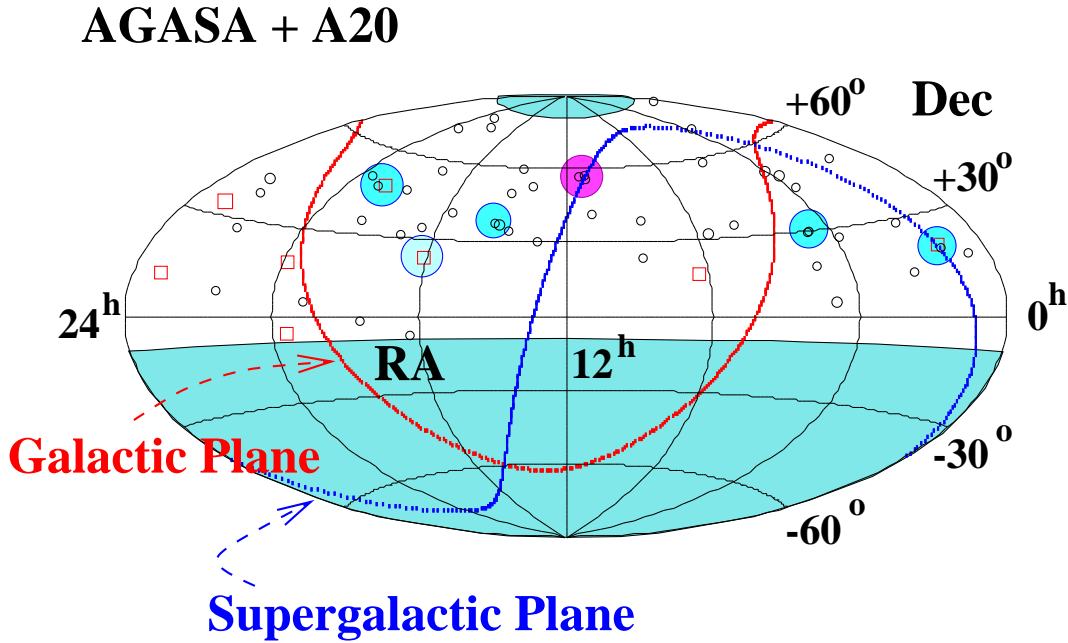}}
\caption{Arrival directions of cosmic rays with energies above 10$^{19.0}$eV 
	on the equatorial coordinates. 
	Open circles, and open squares represent cosmic rays with energies of 
	(4 -- 10) $\times$ 10$^{19}$eV, and $\geq$ 10$^{20}$eV, respectively. 
	The galactic and supergalactic planes are shown by 
	the red and blue curves. 
	Large circles indicate event clusters within 2.5$^{\circ}$ 
	as noted in Table \ref{tbl:40EeV}. 
	\label{fig:equ1960}
}
\end{figure}

Figure \ref{fig:equ1960} shows arrival directions of cosmic rays 
with energies above 4 $\times$ 10$^{19}$eV in the equatorial coordinates. 
Open circles, and Open squares represent cosmic rays 
with energies of (4 -- 10) $\times$ 10$^{19}$eV, 
and $\geq$ 10$^{20}$eV, respectively. 
The shaded regions indicate the celestial regions 
excluded in this paper 
due to the zenith angle cut of $\leq$ 45$^{\circ}$. 
The galactic and supergalactic planes are drawn by the red and blue curves. 
The shaded circle near the center is the C2 triplet 
-- three events are observed within 2.5$^{\circ}$ -- 
and the chance probability of observing such triplet 
under an isotropic distribution is about 1 \%. 
This value becomes somewhat larger than our Ap.J. publication \cite{takeda99a}, 
but this triplet is an interesting phenomenon. 
A new doublet is observed around (14$^h$ 10$^m$, 37.5$^\circ$), 
which is referred to as ``C6'' 
on the basis of our Ap.J. publication \cite{takeda99a}.

\begin{table}[p]
\begin{center}
\caption{AGASA $+$ A20 events above 4 $\times$ 10$^{19}$eV}
\label{tbl:40EeV}

%%\vspace{0.8zh}
\begin{tabular}{lllrrrrc}
 \hline
Date & Time & Energy & \multicolumn{4}{c}{Coordinates} \\
     &      &        & $\alpha$ & $\delta$ & $l^{G}$ & $b^{G}$ \\
 \hline
 84/12/12 & 14:18:02 & 6.81 $\times$ 10$^{19}$ eV & 22$^{h}$ 21$^{m}$ & 38.4$^{\circ}$     &  93.3$^{\circ}$     & $-$15.7$^{\circ}$ &    \\
 84/12/17 & 10:28:16 & 9.79                        & 18$^{h}$ 29$^{m}$ & 35.3$^{\circ}$     &  63.5$^{\circ}$     &  19.4$^{\circ}$ &    \\
 86/01/05 & 19:31:03 & 5.47                        &  4$^{h}$ 38$^{m}$ & 30.1$^{\circ}$     & 170.4$^{\circ}$     & $-$11.2$^{\circ}$ & C4 \\
 86/10/23 & 14:25:15 & 6.22                        & 14$^{h}$ 02$^{m}$ & 49.9$^{\circ}$     &  96.8$^{\circ}$     &  63.4$^{\circ}$ &    \\
 87/11/26 & 17:49:20 & 4.82                        & 21$^{h}$ 57$^{m}$ & 27.6$^{\circ}$     &  82.1$^{\circ}$     & $-$21.1$^{\circ}$ &    \\
     &      &        &          &          &         &         &    \\
 89/03/14 & 02:45:39 & 5.27                        & 13$^{h}$ 48$^{m}$ & 34.7$^{\circ}$     &  68.3$^{\circ}$     &  75.6$^{\circ}$ &    \\
 89/08/16 & 08:32:01 & 4.07                        &  5$^{h}$ 51$^{m}$ & 58.5$^{\circ}$     & 154.5$^{\circ}$     &  15.6$^{\circ}$ &    \\
\hline
 90/11/25 & 11:05:39 & 4.51                        & 16$^{h}$ 17$^{m}$ & $-$7.2$^{\circ}$     &   6.1$^{\circ}$     &  29.6$^{\circ}$ &    \\
 91/04/03 & 00:32:40 & 5.09                        & 15$^{h}$ 47$^{m}$ & 41.0$^{\circ}$     &  65.7$^{\circ}$     &  51.5$^{\circ}$ &    \\
 91/04/20 & 08:24:49 & 4.35                        & 18$^{h}$ 59$^{m}$ & 47.8$^{\circ}$     &  77.9$^{\circ}$     &  18.4$^{\circ}$ & C3 \\
     &      &        &          &          &         &         &    \\
 91/05/31 & 13:07:04 & 5.53                        &  3$^{h}$ 37$^{m}$ & 69.5$^{\circ}$     & 136.6$^{\circ}$     &  11.2$^{\circ}$ &    \\
 91/11/29 & 14:53:03 & 9.10                        & 19$^{h}$ 06$^{m}$ & 77.2$^{\circ}$     & 108.8$^{\circ}$     &  25.6$^{\circ}$ &    \\
 91/12/10 & 18:59:10 & 4.24                        &  0$^{h}$ 12$^{m}$ & 78.6$^{\circ}$     & 121.0$^{\circ}$     &  15.9$^{\circ}$ &    \\
 92/01/07 & 03:16:49 & 4.51                        &  9$^{h}$ 36$^{m}$ & 38.6$^{\circ}$     & 184.3$^{\circ}$     &  48.0$^{\circ}$ &    \\
 92/01/24 & 12:26:17 & 4.88                        & 17$^{h}$ 52$^{m}$ & 47.9$^{\circ}$     &  74.8$^{\circ}$     &  29.4$^{\circ}$ &    \\
     &      &        &          &          &         &         &    \\
 92/02/01 & 17:20:52 & 5.53                        &  0$^{h}$ 34$^{m}$ & 17.7$^{\circ}$     & 117.2$^{\circ}$     & $-$45.0$^{\circ}$ &    \\
 92/03/30 & 03:05:30 & 4.47                        & 17$^{h}$ 03$^{m}$ & 31.4$^{\circ}$     &  53.6$^{\circ}$     &  35.6$^{\circ}$ &    \\
 92/08/01 & 13:00:47 & 5.50                        & 11$^{h}$ 29$^{m}$ & 57.1$^{\circ}$     & 143.2$^{\circ}$     &  56.6$^{\circ}$ & C2 \\
 92/09/13 & 08:59:44 & 9.25                        &  6$^{h}$ 44$^{m}$ & 34.9$^{\circ}$     & 180.5$^{\circ}$     &  13.9$^{\circ}$ &    \\
 93/01/12 & 02:41:13 &  \underline{\bf 10.1}      &  8$^{h}$ 17$^{m}$ & 16.8$^{\circ}$     & 206.7$^{\circ}$     &  26.4$^{\circ}$ &    \\
     &      &        &          &          &         &         &    \\
 93/01/21 & 07:58:06 & 4.46        & 13$^{h}$ 55$^{m}$ & 59.8$^{\circ}$     & 108.8$^{\circ}$     &  55.5$^{\circ}$ &    \\
 93/04/22 & 09:39:56 & 4.42                        &  1$^{h}$ 56$^{m}$ & 29.0$^{\circ}$     & 139.8$^{\circ}$     & $-$31.7$^{\circ}$ &    \\
 93/06/12 & 06:14:27 & 6.49                        &  1$^{h}$ 16$^{m}$ & 50.0$^{\circ}$     & 127.0$^{\circ}$     & $-$12.7$^{\circ}$ &    \\
 93/12/03 & 21:32:47 &  \underline{\bf 21.3}                        &  1$^{h}$ 15$^{m}$ & 21.1$^{\circ}$     & 130.5$^{\circ}$     & $-$41.4$^{\circ}$ & C1 \\
 94/07/06 & 20:34:54 &  \underline{\bf 13.4}$^{\ast}$                        & 18$^{h}$ 45$^{m}$ & 48.3$^{\circ}$     &  77.6$^{\circ}$     &  20.9$^{\circ}$ & C3 \\
     &      &        &          &          &         &         &    \\
 94/07/28 & 08:23:37 & 4.08                         &  4$^{h}$ 56$^{m}$ & 18.0$^{\circ}$     & 182.8$^{\circ}$     & $-$15.5$^{\circ}$ &    \\
 95/01/26 & 03:27:16 & 7.76                        & 11$^{h}$ 14$^{m}$ & 57.6$^{\circ}$     & 145.5$^{\circ}$     &  55.1$^{\circ}$ & C2 \\
 95/03/29 & 06:12:27 & 4.27                        & 17$^{h}$ 37$^{m}$ & $-$1.6$^{\circ}$     &  22.8$^{\circ}$     &  15.7$^{\circ}$ &    \\
 95/04/04 & 23:15:09 & 5.79                        & 12$^{h}$ 52$^{m}$ & 30.6$^{\circ}$     & 117.5$^{\circ}$     &  86.5$^{\circ}$ &    \\
 95/10/29 & 00:32:16 & 5.07                        &  1$^{h}$ 14$^{m}$ & 20.0$^{\circ}$     & 130.2$^{\circ}$     & $-$42.5$^{\circ}$ & C1 \\
 \hline
     &      &        &          &          &         &         &    \\
\multicolumn{8}{l}{\footnotesize
$\ast $     \hspace{1em} The value on the Ap.J. publication is a typo.}\\
\end{tabular}
\end{center}
\end{table}

\begin{table}[p]
\begin{center}
Table 1: AGASA $+$ A20 events above 4 $\times$ 10$^{19}$eV (contd.)

%%\vspace{0.8zh}
\begin{tabular}{lllrrrrc}
 \hline
Date & Time & Energy & \multicolumn{4}{c}{Coordinates} \\
     &      &        & $\alpha$ & $\delta$ & $l^{G}$ & $b^{G}$ \\
 \hline
 95/11/15 & 04:27:45 & 4.89 $\times$ 10$^{19}$ eV  &  4$^{h}$ 41$^{m}$ & 29.9$^{\circ}$     & 171.1$^{\circ}$     & $-$10.8$^{\circ}$ & C4 \\
 96/01/11 & 09:01:21 &  \underline{\bf 14.4}                        & 16$^{h}$ 06$^{m}$ & 23.0$^{\circ}$     &  38.9$^{\circ}$     &  45.8$^{\circ}$ & C5 \\
 96/01/19 & 21:46:12 & 4.80                        &  3$^{h}$ 52$^{m}$ & 27.1$^{\circ}$     & 165.4$^{\circ}$     & $-$20.4$^{\circ}$ &    \\
 96/05/13 & 00:07:48 & 4.78                        & 17$^{h}$ 56$^{m}$ & 74.1$^{\circ}$     & 105.1$^{\circ}$     &  29.8$^{\circ}$ &    \\
 96/10/06 & 13:36:43 & 5.68                        & 13$^{h}$ 18$^{m}$ & 52.9$^{\circ}$     & 113.8$^{\circ}$     &  63.7$^{\circ}$ &    \\
     &      &        &          &          &         &         &    \\
 96/10/22 & 15:24:10 & \underline{\bf 10.5}                        & 19$^{h}$ 54$^{m}$ & 18.7$^{\circ}$     &  56.8$^{\circ}$     &  $-$4.8$^{\circ}$ &    \\
 96/11/12 & 16:58:42 & 7.46                        & 21$^{h}$ 37$^{m}$ &  8.1$^{\circ}$     &  62.7$^{\circ}$     & $-$31.3$^{\circ}$ &    \\
 96/12/08 & 12:08:39 & 4.30                        & 16$^{h}$ 31$^{m}$ & 34.6$^{\circ}$     &  56.2$^{\circ}$     &  42.8$^{\circ}$ &    \\
 96/12/24 & 07:36:36 & 4.97                        & 14$^{h}$ 17$^{m}$ & 37.7$^{\circ}$     &  68.5$^{\circ}$     &  69.1$^{\circ}$ & C6   \\
 97/03/03 & 07:17:44 & 4.39                        & 19$^{h}$ 37$^{m}$ & 71.1$^{\circ}$     & 103.0$^{\circ}$     &  21.9$^{\circ}$ &    \\
     &      &        &          &          &         &         &    \\
 97/03/30 & 07:58:21 &  \underline{\bf 15.0}                        & 19$^{h}$ 38$^{m}$ & $-$5.8$^{\circ}$     &  33.1$^{\circ}$     & $-$13.1$^{\circ}$ &    \\
 97/04/28 & 13:46:18 & 4.20                        &  2$^{h}$ 18$^{m}$ & 13.8$^{\circ}$     & 152.9$^{\circ}$     & $-$43.9$^{\circ}$ &    \\
 97/11/20 & 07:23:25 & 7.21                        & 11$^{h}$ 09$^{m}$ & 41.8$^{\circ}$     & 171.2$^{\circ}$     &  64.6$^{\circ}$ &    \\
 98/02/06 & 00:12:26 & 4.11                        &  9$^{h}$ 47$^{m}$ & 23.7$^{\circ}$     & 207.2$^{\circ}$     &  48.6$^{\circ}$ &    \\
 98/03/30 & 08:17:26 & 6.93                        & 17$^{h}$ 16$^{m}$ & 56.3$^{\circ}$     &  84.5$^{\circ}$     &  35.3$^{\circ}$ &    \\
     &      &        &          &          &         &         &    \\
 98/04/04 & 20:07:03 & 5.35                        & 11$^{h}$ 13$^{m}$ & 56.0$^{\circ}$     & 147.5$^{\circ}$     &  56.2$^{\circ}$ & C2 \\
 98/06/12 & 06:43:49 &  \underline{\bf 12.0}                        & 23$^{h}$ 16$^{m}$ & 12.3$^{\circ}$     &  89.5$^{\circ}$     & $-$44.3$^{\circ}$ &     \\
 98/09/03 & 23:42:48 & 4.69                        & 19$^{h}$ 36$^{m}$ & 50.7$^{\circ}$     &  83.1$^{\circ}$     &  14.0$^{\circ}$ \\
 98/10/27 & 00:45:37 & 6.11                        &  3$^{h}$ 45$^{m}$ & 44.9$^{\circ}$     & 152.4$^{\circ}$     &  $-$7.8$^{\circ}$ \\
 99/01/22 & 08:43:54 & 7.53                        & 19$^{h}$ 11$^{m}$ &  5.3$^{\circ}$     &  39.9$^{\circ}$     &  $-$2.1$^{\circ}$ \\
     &      &        &          &          &         &         &    \\
 99/07/22 & 08:11:14 & 4.09                        &  7$^{h}$ 39$^{m}$ & 32.2$^{\circ}$     & 187.5$^{\circ}$     &  23.6$^{\circ}$ \\
 99/07/28 & 04:08:49 & 7.16                        &  3$^{h}$ 46$^{m}$ & 49.5$^{\circ}$     & 149.8$^{\circ}$     &  $-$4.0$^{\circ}$ \\
 99/09/22 & 01:43:30 &  \underline{\bf 10.4}                        & 23$^{h}$ 03$^{m}$ & 33.9$^{\circ}$     &  98.5$^{\circ}$     & $-$23.8$^{\circ}$ \\
 99/09/25 & 20:13:49 & 4.95                        & 22$^{h}$ 40$^{m}$ & 42.6$^{\circ}$     &  98.8$^{\circ}$     & $-$14.0$^{\circ}$ \\
 99/10/20 & 03:46:21 & 6.19                        &  4$^{h}$ 37$^{m}$ &  5.1$^{\circ}$     & 191.3$^{\circ}$     & $-$26.5$^{\circ}$ \\
     &      &        &          &          &         &         &    \\
 99/10/20 & 22:01:35 & 4.29                        &  4$^{h}$ 02$^{m}$ & 51.7$^{\circ}$     & 150.3$^{\circ}$     &  $-$0.7$^{\circ}$ \\
 00/05/26 & 18:38:16 & 4.98                        & 14$^{h}$ 08$^{m}$ & 37.1$^{\circ}$     &  69.3$^{\circ}$     &  71.0$^{\circ}$ & C6 \\
 \hline
 97/04/10 & 02:48:48 & 3.89$^{\ast \ast}$   & 15$^{h}$ 58$^{m}$ & 23.7$^{\circ}$     &  39.1$^{\circ}$     &  47.8$^{\circ}$ & C5 \\
 \hline
     &      &        &          &          &         &         &    \\
\multicolumn{8}{l}{\footnotesize
$\ast \ast$ \hspace{1em} This energy is just below the 4 $\times$ 10$^{19}$eV cut.
} \\
\end{tabular}
\end{center}
\end{table}

%*********************************************************************
\section*{Acknowledgments}

We are grateful to Akeno-mura, Nirasaki-shi, Sudama-cho, Nagasaka-cho, 
Ohizumi-mura, Tokyo Electric Power Co. and 
Nihon Telegram and Telephone Co. for their kind cooperation. 
The authors are indebted to other members of the Akeno group 
in the maintenance of the AGASA array.

%*********************************************************************
%	References

\end{document}